\documentstyle[fleqn]{article}
\begin{document}

%                          Title
\begin{center}
{\large \bf RENORMALIZATION WITHOUT REGULARIZATION AND R-OPERATION}
\\

%                      author/address
\vspace{4mm}
Fyodor A. Lunev \\
 Department of Physics, Moscow State University, Moscow, 119899,
Russia

\end{center}

\vspace{4mm}

%                       Abstract

Definition of Feynman integrals as solutions of some well defined
systems of differential equations is proposed.
 This definition is equivalent to usual
one but needs no regularization and application of $R$-operation.
It is argued that proposed renormalization scheme maintains all
symmetries that can be maintained in perturbative quantum field
theory,
and also is convenient for practical calculations.

\vspace{16mm}

%                         Text

{\bf 1.Motivation.}

 \vspace{2mm}

Though  fundamental   results  in   renormalization
theory were obtained    many    years    ago    in    classical
works   of Feynman,   Tomonaga,   Schwinger,    Dyson,   Salam,
Bogolubov, Parasiuk,  Hepp,   and   Zimmermann\footnote{Beautiful
account of foundations and modern achievements of renormalization
theory can be found, for instance, in monographs \cite{Cl}},
renormalization problems continue to attract the attention of
theorists.   In particular, during last twenty years   very many
papers were devoted to investigations of various renormalization
schemes.

Of course,  all known  renormalization schemes  are equivalent,  in
principal, at perturbative level. However, their practical  value
is different. To be useful, the renormalization scheme must

\begin{itemize}
\item  maintain symmetries
\item be convenient for practical calculations
\end{itemize}

 Usually, it is hard to satisfy both this conditions.
Indeed, for application of standard renormalization schemes one must

\begin{itemize}
\item fix some regularization
\item apply R-operation with some fixed parameters
\end{itemize}

\noindent But

\begin{itemize}
\item symmetries can be broken at both stages
\item regularization usually hampers the evaluation of diagrams
\end{itemize}

 So, I believe, it is interesting to formulate renormalization
scheme that needs no any regularization and application of R-
operation
at all.

Such renormalization  scheme is presented  in  this  report. It  is
equivalent to usual R-operation  scheme. But "equivalent"
doesn't mean "the same". Indeed, in standard  R-operation
renormalization  scheme  one  must,  first,  regularize  initial
divergent (in  general) Feynman  integral.Then it is necessary
to use rather complicated subtraction  prescription (forest
formula) to obtain finite   result.    Nothing    similar   is
needed in    my renormalization  scheme.  To  obtain  finite
expression for given Feynman  integral,  one  must   only  solve
some well   defined differential  equations.  Neither  any
regularization,  nor  any manipulations  with  counterterm
diagrams  are  needed to obtain finite result.

\vspace{4mm}

{\bf 2. Differential equations for definition and evaluation of
Feynman
integrals.}

\vspace{2mm}

{\em 2.A. The simplest example.}

\vspace{2mm}

To clarify the main idea
of presented work, let us consider the simplest divergent
(Euclidean) Feynman diagram (see Fig.1). For simplicity, we will
consider massless case.  Then in coordinate space this diagram is
well
defined function

\begin{figure}[h]
%\senterline{\psfig{figure=ptxda.eps,width=8cm}
\unitlength=5mm
\centering
\thicklines
\begin{picture}(8,7)
\put(1,4){\line(1,0){1}}
\put(4,4){\oval(4,2)}
\put(6,4){\line(1,0){1}}
\end{picture}
\caption{The simplest divergent Feynman diagram}
\end{figure}
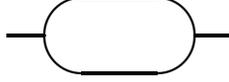

\begin{equation}
\tilde {\Gamma} (x) = \frac{(2\pi)^4}{(x^2)^2} \label{1}
\end{equation}

But the function $\tilde {\Gamma} (x) $ is not a distribution. So its
Fourier image ${\Gamma} (p)$ doesn't exist. But $\tilde {\Gamma} (x)$
is defined as distribution on the set

\begin{equation}
S_0 = \{\phi (x) \in S: \ \phi (0) =0 \}
\end{equation}

\noindent where $S$ is Schwartz space. So the problem of
renormalization theory can be formulated as follows:

\begin{quote}
{\em to define
$\tilde {\Gamma}^{ren}(x) \in S'$ in such a way that}
\end{quote}

\begin{equation}
\tilde {\Gamma}^{ren}
\raisebox{-1.5em}{\vrule height 1.5em \ $S_0$}
=\tilde {\Gamma}  \label{2}
\end{equation}

One notes that the distribution  $\tilde {\Gamma}^{ren}(x)$, defined
as
a solution of equation

\begin{equation}
x^2 \tilde {\Gamma}^{ren}(x) = \frac{(2 \pi)^4}{x^2},   \label{3}
\end{equation}

\noindent satisfies the conditions formulated above. (This follows
from
Hormander's theorem about possibility of dividing by polynomial in
$S'$). So we see that it is very natural to define ${\Gamma}^{ren}(p)
$
as a solution of the differential equation

\begin{equation}
\Delta  {\Gamma}^{ren}(p) = - \frac{4 {\pi}^2}{p^2} \label{4}
\end{equation}

\noindent that is equivalent to (\ref{3}). The general solution of
equation (\ref{4}) from $S'$ is given by formula

\begin{equation}
{\Gamma}^{ren} = - \pi ^2 \ln \frac{p^2}{\mu ^2} \label{5}
\end{equation}

\noindent where $\mu ^2$ is the constant of integration. This
expression coincide with usual one for "bubble" diagram in Fig.1. We
see that it is possible to obtain the standard result for diagram
under
consideration without using of any regularization and any subtraction
procedure.

\vspace{2mm}

{\em 2.B. General formalism}

\vspace{2mm}

Without loss of generality, we will consider only 1PI
(Euclidean) diagram $\Gamma _L$ without internal vertexes. In
coordinate space

\equation
{\tilde {\Gamma}}_L (x_1,...,x_n; \{ m_{ij}^2 \}) = \prod_{all \
lines \ of \ \Gamma} D(x_i - x_j; m_{ij}^2) \label{6}
\endequation

\noindent where $L$ is the number of loops and $m^2_{ij}$ are masses
in
propagators.

Below we will interpret $ m_{ij}^2 $ as the square of some
Euclidean two dimensional vector. Then we can write

\begin{equation}m_{ij}^2 = (m_{ij,1})^2 + (m_{ij,2})^2 \end{equation}

\noindent Further, let us define

\begin{equation}\hat D (x,u) = \int \mbox{d}^2 m e^{i \vec u \vec m}
D(x,m)=
\int \mbox{d}^4 p \mbox{d}^2 m \frac{e^{ipx + i \vec u \vec
m}}{p^2 +m^2} = \frac{16 \pi ^3}{(x^2 + m^2)^2}
\end{equation}

\noindent and

\begin{eqnarray}
 \hat {\Gamma} _n (x_1,...,x_n; \{ u_{ij}^2 \}) \equiv
\int \prod _{all \ lines \ of \ \Gamma } \mbox{d} ^2 m_{ij} \exp
\left(
i \sum _{all \ lines \ of \ \Gamma } \vec m _{ij} \vec u _{ij}
\right)
\tilde \Gamma _n (x_1,...,x_n; \{m_{ij}^2 \} ) \nonumber \\
=(16 \pi ^3)^N \frac{1}{P(x_1,...,x_n; \{ u_{ij}^2 \} ) }
\end{eqnarray}

\noindent where $N$ is total number of lines in diagram $\Gamma$
and $P$ is the polynomial:
\equation
  P = \prod _{all \ lines \ of \ \Gamma } [(x_i - x_j)^2 +
u_{ij}^2]^2
\endequation

One can see that ${\hat \Gamma}_L \not \in S'$, but ${\hat \Gamma}_L$
is defined as distribution on the space of test functions

\begin{equation} S_0 = \{ \phi \in S': \ \
\phi
\raisebox{-1.5em}{\vrule height 1.5em \ $x_i =x_j$}
=0,  \ \
\partial _{\alpha}
\phi
\raisebox{-1.5em}{\vrule height 1.5em \ $x_i =x_j$}
=0, \ ... \}
\end{equation}

\noindent {\em The function $ \hat {\Gamma}^{ren}_L $ can be defined
as
the prolongation of $ \hat {\Gamma}_L $ on the whole space $S$.}

Now we can give the following inductive definition of
$ \hat {\Gamma}^{ren}_L $.

Let $ \hat {\Gamma}^{ren}_{L-1} $ is already defined. Then
$ \hat {\Gamma}^{ren}_L $ is defined by the following equations in
coordinate space:

\begin{equation}
[(x_i - x_j)^2 + u^2_{ij}]^2  \hat {\Gamma}^{ren}_L (x_1,...,x_n,
 \{ u^2_{kl} \} )
 =
(16 \pi ^3)  \hat {\Gamma}^{ren}_{L - 1} (x_1,...,x_n, \{ u^2_{kl} \}
)
\raisebox{-1.5em}{\vrule height 1.5em \
$(kl) \not = (ij) $}
\end{equation}

\noindent or, if

\begin{equation}
\bigtriangleup _{(p_i -p_j)m_{ij}} = \sum _{\alpha =1}^{4} \left(
\frac {\partial}{\partial p_{i,\alpha}} - \frac {\partial}{\partial
p_{j, \alpha}} \right)^2 + \sum _{\alpha =1}^{2} \frac {
{\partial}^2}{{\partial m_{ij, \alpha}}^2} \end{equation}

\noindent by the following equivalent equations in the momentum space:

\begin{equation}
[ \Delta _{(p_i - p_j)m_{ij}} ]^2
{\Gamma}^{ren}_{L} (p_1,...,p_n, \{ m^2_{kl} \} )
= 16 \pi ^3 \delta ( \vec {m_{ij}} )
{\Gamma}^{ren}_{L-1} (p_1,...,p_n, \{ m^2_{kl} \} )
\raisebox{-1.5em}{\vrule height 1.5em \
$(kl) \not = (ij) $}
\label{7}
\end{equation}

\noindent with boundary conditions

\equation
\lim_{|p | \rightarrow \infty}  \frac {1}{p^{\omega (\Gamma ) +
\epsilon}} \Gamma _L^{ren} = 0; \ \
\lim_{m_{ij}  \rightarrow \infty}  \frac {1}{m^{\omega (\Gamma
) + \epsilon}_{ij}} \Gamma _L^{ren}= 0.  \label{8}
\endequation

One can prove that $ \Gamma _L^{ren}$ is defined by equations (\ref 7
)
with boundary conditions (\ref 8 ) up to polynomial of the degree $
\omega (\Gamma _L) $ with respect to $p_1,  ...,p_n$, just as in
usual
formulation of the renormalization theory.

In general, the following theorem of equivalence is valid:

\begin{quote}
{\em Let $ \Gamma _L^{\Lambda}$ is the diagram regularized by cutoff
at
$|p_i| = \Lambda$, and  $ \Gamma _L^{ren}$ is obtained from
$ \Gamma _L^{\Lambda}$ by means of usual R-operation. Then
$ \Gamma _L^{ren}$ satisfies equations} (\ref 7).
\end{quote}

\noindent The proof of this theorem is given in \cite{L}.

\vspace{2mm}

{\em 2.C. Symmetries.}

\vspace{2mm}

In quantum field theory any symmetry corresponds to certain Ward
 identity, and to maintain the  symmetry in given renormalization
scheme means to fix the freedom in definition of divergent Feynman
integrals in such a way that the corresponding Ward identities are
satisfied.

But in our renormalization scheme  arbitrariness in definition of
divergent Feynman diagram is not fixed {\em a priory}.
 So, if there exists the renormalization scheme with counterterms
that are polynomial with respect to masses and compatible with Ward
identities, that, due to equivalence theorem, \footnote{Strictly
speaking, for applicability of equivalence theorem the
given renormalization scheme must be equivalent to cutoff one up to
finite renormalization. To author's knowledge, all known
renormalization schemes  satisfies this condition.}
formulated above, the equations (\ref 7) are also compatible with
Ward
identities.  So we can state that {\em proposed renormalization
scheme
maintains all symmetries that can be conserved in perturbative
quantum
field theory}.

\vspace{2mm}

{\em 2.D. Application to evaluation of concrete diagrams.}

\vspace{2mm}

For the lack of the place, we cannot give here example of non-trivial
application of proposed renormalization scheme to evaluation of
concrete Feynman diagrams. So we give only some notes concerning
possible advantages of our approach in comparison with dimensional
regularization scheme that now is the most popular one in practical
calculations.

Feynman integrals are  rather complicated multiple ones.
When one evaluates such integrals in non-integer dimension, then,
after
the first integration one obtains usually rather complicated non
elementary function that hampers further integration. In our
approach,
one always works in four dimensions with mathematically well defined
objects that need no any regularization. This allows sometimes to
carry
out exact evaluation of Feynman diagrams in four dimensions in
situations when exact evaluation in noninteger dimensions is
impossible. The example of such calculations in the case of two loop
self-energy diagram with three propagators is  given in (\cite{LL}).

\vspace{4mm}

{\bf 3. Second order equations for definition of Feynman integrals.}

\vspace{2mm}

In our approach divergent as well as convergent Feynman diagrams are
defined by equations (\ref 7) with boundary conditions (\ref 8).
There
exists also other possibilities for definition of Feynman integrals
as
the solutions of the systems of differential equations. For instance,
if one  considers $m^2$ as the square of the {\em four}
dimensional vector (rather than two dimensional as above), then, at
least in the case $\omega (\Gamma _L) <2$, one can obtain the
following
equations:

\begin{equation}
\bigtriangleup _{(p_i -p_j)m_{ij}}
\left( \frac{\Gamma _L}{m^2_{ij}} \right) =
 \/ - (2 \pi)^2 \delta (\vec m )
 \Gamma _L
\raisebox{-1.5em}{\vrule height 1.5em  $m_{ij}=0$}
 \label{9}
\end{equation}

\noindent whereas the function $
 \Gamma _L
\raisebox{-1.5em}{\vrule height 1.5em  $m_{ij}=0$}$
satisfies the equations

\begin{equation}
\Delta _{p_i-p_j}
 \Gamma _L
\raisebox{-1.5em}{\vrule height 1.5em  $m_{ij}=0$}
= - \frac{1}{(2 \pi)^2} \Gamma_{L-1}
\label{9a}
\end{equation}

\noindent (In RHS of equation(\ref {9a}) $\Gamma _{L-1}$ means the
initial diagrams $\Gamma _{L}$ without the propagator with the mass
$m_{ij})$.

The equation (\ref 9) and (\ref{9a}) are the second order ones
whereas
equations (\ref 7) are of the forth order. This fact simplifies the
investigation of the corresponding Feynman integrals. In particular,
one can easy to obtain from (\ref 9) the formula that connects
Feynman
diagrams with massive and massless lines:

\begin{equation}
\Gamma _L = \frac{m^2_{ij}}{\pi ^2} \int d^4 p'
\frac{1 }{[(p-p')^2 +m_{ij}^2]^3}
\Gamma _L
(p',...)
\raisebox{-1.5em}{\vrule height 1.5em  $m_{ij}=0$}
\end{equation}

There exist also many other equivalent definitions of divergent
Feynman
integrals by means of second order systems of differential equations.
All of them can be obtained by means of various modifications of
methods presented in section 2.

\vspace{4mm}

{\bf 4.Conclusion.}

\vspace{2mm}

Now it is unclear, whether our approach to renormalization theory
has principal advantages in comparison with standard
formulation. But, at least, calculations, presented in \cite{L,LL},
show that our approach gives new effective methods of evaluation
Feynman integrals. So author believes that proposed formalism
will be useful in various investigations in quantum field
theory.

Author is indebted to D.J.Broadhurst, A.I.Davydychev and
V.A.Smirnov for valuable discussions and comments.

%            REFERENCES

\end{document}